\documentclass{svproc}

\usepackage{graphicx}
\usepackage{amsmath}
\usepackage{amsfonts}

\newcommand{\specialcell}[2][c]{%
\begin{tabular}[#1]{@{}c@{}}#2\end{tabular}}

\usepackage{url}

\usepackage{hyperref}

\begin{document}
\mainmatter              
\title{An Intrinsic Framework of Information Retrieval Evaluation Measures}
\titlerunning{An Intrinsic Framework}  
%
\author{Fernando Giner}
\authorrunning{F. Giner} 
%
%
\institute{E.T.S.I. Inform\'{a}tica UNED, \\ C/ Juan del Rosal, 16, 28040-Madrid, Spain,\\
\email{fginer3@gmail.com} \\ 
ORCiD: 0000-0002-9161-0458}

\maketitle              

\begin{abstract}
Information retrieval (IR) evaluation measures are cornerstones for determining the suitability and task performance efficiency of retrieval systems. Their metric and scale properties enable to compare one system against another to establish differences or similarities. Based on the representational theory of measurement, this paper determines these properties by exploiting the information contained in a retrieval measure itself. It establishes the \emph{intrinsic framework} of a retrieval measure, which is the common scenario when the domain set is not explicitly specified. A method to determine the metric and scale properties of any retrieval measure is provided, requiring knowledge of only some of its attained values. The method establishes three main categories of retrieval measures according to their intrinsic properties. Some common user-oriented and system-oriented evaluation measures are classified according to the presented taxonomy.
\end{abstract}
\section{Introduction}
\label{sec:introduction}
Information Retrieval (IR) is a field dealing with the analysis, storage and searching of a user's information need~\cite{salton1968automatic}. To effectively compare and progressively develop better IR systems, assessment plays an important role. Even being evaluation a rich scenario that has contributed to the development of the IR field, a better comprehension of evaluation measures is needed. Specifically, their classification according to the scale types of Stevens~\cite{stevens1946theory}, which helps to make explicit the assumptions behind measures, and to justify the validity of conclusions as consequence of the obtained results.

Due to the empirical nature of IR, this task is not exempt of the historic confusion and controversy about the scale types and the statistical methods that can be performed~\cite{hand1996statistics,michell1986measurement,velleman1993nominal}. Recently, two approaches have discussed the role of IR evaluation metrics\footnote{Here, the commonly used term ``IR evaluation \emph{metric}'' collides with the mathematical term ``\emph{metric}'', which will be used later in this paper. To solve this issue, the rest of the paper will refer the term ``IR evaluation \emph{metrics}'' as ``IR evaluation \emph{measures}'', keeping the term ``\emph{metric}'' for its mathematical sense.} as being interval scales. On the one hand, Moffat~\cite{moffat2022batch} states that retrieval measures can be considered interval scales, whenever they have a real-world basis (an external validity) and can be motivated as corresponding to the usefulness of system outputs. Advocated with rhetoric arguments, this viewpoint shows similarities to the \emph{operational paradigm}~\cite{michell1986measurement,michell2014introduction}. In contrast, Ferrante et al.~\cite{ferrante2021towards,ferrante2022response} investigate the implications of retrieval measures being interval scales, by grounding their arguments~\cite{ferrante2018general} on the \emph{representational theory of measurement} (RTM)~\cite{roberts1985measurement,krantz1971foundations,krantz1989foundations,luce1990foundations}. Regarding this viewpoint, which could be termed as the \emph{representational paradigm}~\cite{michell1986measurement,michell2014introduction}, interval scales are real mappings whose attained values are equispaced. This position seems to have a sound theoretical basis; however, the recent arguments provided~\cite{ferrante2018general} have a limited application~\cite{giner2023comment}. 

The goal of this paper is not to challenge or agree with any of these two viewpoints, neither to find common points shared by both. In fact, they consider different measurement theories since the controversy lies in the assumption or not of the RTM, which is not mentioned in Moffat's work. The goal of this paper is to exemplify the fact of considering retrieval measures as equispaced mappings~\cite{ferrante2021towards} by providing a classification of IR evaluation measures, and to give some insights about the arguments of this position by overcoming the concerns of~\cite{giner2023comment}.

Retrieval measures can be classified based on several factors, such as the domain, range, relation, or expression among others. Probably, the analytical expression is one of the main aspects that has been considered, since it determines the performance of retrieval systems~\cite{sanderson2010test} or differentiates user browsing models~\cite{carterette2011system,moffat2017incorporating}. Along with the analytical expression, the relationship between the items of the domain set and the range set also accounts for the type of a retrieval measure. For instance, the metric properties of the domain set detect whether every item to be measured can be distinguished~\cite{frechet1906quelques}, or the relationships on the range set determine the operations that can be performed, and the statistics that can be applied~\cite{stevens1946theory}. Based on the RTM, this paper shows that the attained values of a retrieval measure determine its metric and scale properties. It establishes the \emph{intrinsic framework}, which exploits the information contained in the retrieval measure itself to provide a taxonomy of IR evaluation measures.

The intrinsic framework is not limited to the IR field. In the area of databases or data mining among others, many empirical studies use the same or similar measures presented here. In studies that are relied on the RTM, the scale characterization is central to make statistical inferences regarding an attribute to be measured. The metric characterization is also useful to determine whether the distance associated to the measure is similar to the Euclidean geometry, or it is a more elastic geometry, where the notion of ``closeness'' allows the existence of distinct points whose distance is zero. At theoretical level, the intrinsic framework provides a formal basis that every retrieval measure assumes when its empirical domain is not explicitly specified. This paper describes some consequences about retrieval measures under the assumption of the RTM, which can be useful for IR theory works that challenge or agree with the representational paradigm.

The rest of the paper is organised as follows: Section~\ref{sec:SoA} reports some related work. In Sections~\ref{sec:formalization} and~\ref{sec:intrinsic-prop}, the intrinsic framework is presented, and its metric and scale properties are characterised. Section~\ref{sec:intrinsic-categories} provides a taxonomy to determine the intrinsic category of a retrieval measure. In Section~\ref{sec:examples}, some common IR evaluation measures are classified with the intrinsic framework. Finally, in Section~\ref{sec:conclusion}, some conclusions are drawn.

\section{Related Work}
\label{sec:SoA}
In the IR field, there is a great body of research work on evaluation~\cite{harman2011information,robertson2008history}, which has lead to consider it as a key area~\cite{allan2003challenges}. Large-scale campaigns and initiatives, such as TREC~\cite{voorhees2005trec}, NTCIR~\cite{sakai2021evaluating}, CLEF~\cite{ferro2019information}, FIRE and INEX, have promoted improvements in academy and industry. The Cranfield 2 experiments~\cite{cleverdon1991significance}, are considered the first attempt of evaluation in the IR field~\cite{croft2010search}, and is the underlying framework of many modern experiments. IR systems are usually compared with a set of topics or search requests \cite{carmel2010estimating}, where IR evaluation measures are computed~\cite{voorhees2006trec,voorhees2003overview}. The attained values of IR evaluation measures are commonly supported with significance test results or confidence intervals~\cite{carterette2012multiple,hull1993using,sakai2014statistical,savoy1997statistical,urbano2019statistical}.  Empirically quantifying and statistically assessing the performance of IR systems enable to establish differences and similarities~\cite{sakai2013metrics}. 

The formal analysis of IR evaluation measures has contributed to a better understanding of their meaning. Some works have shown that retrieval measures correspond to different user browsing models~\cite{azzopardi2018measuring,carterette2011system,chapelle2009expected,moffat2017incorporating,wicaksono2020metrics,zhang2017evaluating}. Others have characterised the effectiveness of retrieval measures with formal properties~\cite{amigo2009comparison,amigo2013general,huibers1996axiomatic,moffat2013seven,sebastiani2015axiomatically,swets1963information}, which help to know the appropriateness of retrieval measures on a specific scenario. The use of formal properties as a method to explore retrieval models and how best to improve them, in order to achieve higher retrieval effectiveness has been fostered by Fang et al.~\cite{fang2004formal,fang2011diagnostic,fang2005exploration,fang2006semantic}, and successfully applied to the study of basic models~\cite{fang2007axiomatic,fang2004formal}, pseudo-relevance feedback methods~\cite{clinchant2011document,clinchant2013theoretical,montazeralghaem2016axiomatic}, translation retrieval models~\cite{karimzadehgan2012axiomatic,rahimi2020axiomatic} and neural network retrieval models~\cite{rosset2019axiomatic}.

There have been several approaches, which explore retrieval measures from a measurement viewpoint, van Rijsbergen~\cite{van1979information,van1974foundation} tackled the foundations of measurement in IR through a conjoint (additive) structure based on precision and recall, then he examined the properties of a measure on this prec-recall structure. Bollman et al.~\cite{bollmann1980measurement} defined a similar conjoint structure, but on the contingency table of the binary retrieval; then, they studied the properties of the proposed MZ-metric. Later, Bollman~\cite{bollmann1984two} shown that retrieval measures can be expressed as a linear combination of the number of relevant\slash nonrelevant retrieved documents, whenever they satisfy two proposed axioms. Flach~\cite{flach2019performance} modelled the empirical domain through  confusion matrices, then the relation between measurement theory and machine learning evaluation is sketched. However, these works do not address the scale properties of retrieval measures.

Some works consider the gold standard as measurement, they analyse the scale properties in ordinal classification~\cite{baccianella2009evaluation,gaudette2009evaluation,vanbelle2009note}, or the scale properties of the ground truth, human annotation or predicted variables~\cite{han2019transforming}. Other works consider the system outputs and the gold standard as independent measurements. It allows to introduce axioms over the similarity of assessors scales and system scales~\cite{busin2013axiometrics,maddalena2014axiometrics}; and to provide a single and unified explanation for most classification, ranking, and clustering measures~\cite{amigo2020nature}, or a methodology to determine the most appropriate task/metric formalization for a given data mining problem~\cite{amigo2021my}.

Recently, Fuhr~\cite{fuhr2018some} proposed some experimental protocols to measure the usefulness of IR systems, such as avoiding the use of MRR and ERR since they violate basic requirements for a metric. In contrast, Sakai~\cite{sakai2021fuhr} argues some of Fuhr's statements since they do not explain the experimental alignment between retrieval measures and user's perception of usefulness. As a consequence or parallel to this difference, there exists a currently active dialogue~\cite{ferrante2021towards,moffat2022batch,ferrante2022response}, which considers retrieval measures as real mappings that quantify the usefulness of retrieval systems. Based on the RTM, Ferrante et al.~\cite{ferrante2021towards,ferrante2022response} investigate the implications of IR measures being interval scales. This viewpoint claims that interval scales are real mappings whose attained values are equispaced. They propose intervalization as a feasible technique to obtain meaningfulness. In contrast, Moffat~\cite{moffat2022batch} considers that document rankings are categorical data, retrieval measures are numeric mappings defined by the context of the dataset, and they are bounded to a set of target values by some external reality. Moffat claims that retrieval measures can be considered interval scales, whenever they have a real-world basis, i.e., an external validity; for instance, a prize assignation to classes of rankings. 

These two viewpoints determine the scale type of retrieval measures based on different assumptions. The former, which could be termed the \emph{representational paradigm}, grounds its arguments on the work of Ferrante et al.~\cite{ferrante2018general}. They developed a framework for both set-based and rank-based IR evaluation measures as well as both  binary and multi-graded relevance, determining whether retrieval measures are interval scales. However, the provided arguments have a limited application~\cite{giner2023comment}, which can be addressed by allowing the domain set to be specified by the retrieval measure itself. The intrinsic framework presented here follows this point of view. 

\section{Formalisation of the Intrinsic Framework}
\label{sec:formalization}
In \emph{batch evaluation}, a topic or query is submitted to an IR system, which returns as output a \emph{search engine result page} (SERP)\footnote{Typically a SERP includes content in a non homogeneous manner, such as images, query suggestions, knowledge panels, etc. However, here, we consider the classical ordered (or unordered) list of documents since it is the common structure considered when the evaluation of ranking models is studied.}, $\mathbf{\hat{r}}$. Then, evaluation measures quantify, in numeric terms, the \emph{effectiveness} of the retrieval system, i.e., its ability to leave aside nonrelevant documents while retrieving relevant ones. Thus, an effectiveness measure can be seen as a mapping that relates a set of possible rankings, $\mathbf{R}$, with real numbers, i.e., an IR evaluation measure assigns numbers (numerical range) that correspond to a set of rankings (empirical domain). Once a retrieval measure has been defined, it has consequences on both the empirical domain and the numerical range. 

\begin{figure}
  \centering
  \scalebox{0.25}
  {
  \includegraphics[width=\linewidth]{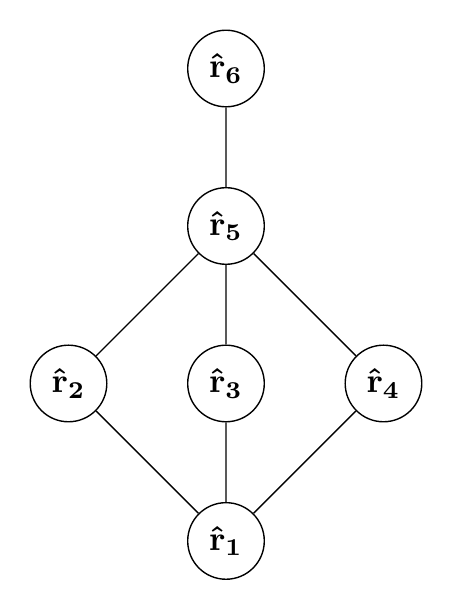}
  }
  \caption{Example of Hasse diagram, $G_{f}$, associated to a retrieval measure.}
	\label{fig:hasse-diagram}
\end{figure}

On the empirical domain, every retrieval measure, $f$, establishes an implicit ordering relationship, $\preceq_{f}$, which is defined by their attained values as follows: 
\begin{equation}
\label{eq:associted-ordering}
\mathbf{\hat{r}_1} \preceq_{f} \mathbf{\hat{r}_2} \Longleftrightarrow f(\mathbf{\hat{r}_1}) \leq f(\mathbf{\hat{r}_2}) \ ,
\end{equation}
for all $\mathbf{\hat{r}_1}$, $\mathbf{\hat{r}_2} \in \mathbf{R}$; therefore, every IR evaluation measure has a naturally associated ordering, which is inherently derived from the measure itself. Every pair of elements is comparable with this binary relationship, $\preceq_{f}$, and the transitivity is verified trivially; thus, it is a \emph{weak order}\footnote{The associated weak order, $\preceq_f$, may be transformed into a total order by considering the following equivalence relation: $\mathbf{\hat{r}_1} \sim_{f} \mathbf{\hat{r}_2} \Leftrightarrow f(\mathbf{\hat{r}_1}) = f(\mathbf{\hat{r}_2})$. Let $\mathbf{R^{*}}$ be the set of equivalence classes, and let $\mathbf{\hat{r}^{*}_1}$ and $\mathbf{\hat{r}^{*}_2}$ be two elements of this set containing the individual system output rankings $\mathbf{\hat{r}_1}$, $\mathbf{\hat{r}_2} \in \mathbf{R}$, respectively. It can be defined the following ordering on $\mathbf{R^{*}}$: $\mathbf{\hat{r}^{*}_1} \preceq_{f}^{*} \mathbf{\hat{r}^{*}_2} \Leftrightarrow \mathbf{\hat{r}_1} \preceq_{f} \mathbf{\hat{r}_2}$. Then, $(\mathbf{R^{*}}, \preceq_{f}^{*})$ is called the \emph{reduction} or \emph{quotient} of $(\mathbf{R}, \preceq_{f})$, where $\preceq_{f}^{*}$ is well-defined and $(\mathbf{R^{*}}, \preceq_{f}^{*})$ is a totally ordered set~\cite{roberts1985measurement}.}~\cite{roberts1985measurement}. The ordering structure, $(\mathbf{R}, \preceq_f)$, can be represented with a Hasse diagram, denoted by $G_{f}$, where nodes are labelled with the elements of $\mathbf{R}$. They are placed in different levels depending on the attained values of $f$. An edge indicates that the attained value of one element is greater than the other and there are no other rankings between them. For instance, the associated graph of a retrieval measure, $f$, such that $f(\mathbf{\hat{r}_1}) < f(\mathbf{\hat{r}_2})=f(\mathbf{\hat{r}_3})=f(\mathbf{r_4})<f(\mathbf{\hat{r}_5})< f(\mathbf{\hat{r}_6})$, is shown in Fig.~\ref{fig:hasse-diagram}.

On the numerical range, the attained values of an IR evaluation measure enable to examine one system against another to establish relationships, differences or similarities. The effectiveness of a pair of retrieval systems is usually compared by considering the absolute/relative difference of the attained values or with testing validation based on this difference~\cite{sanderson2010test}. Thus, in order to perform comparisons, any retrieval measure has a naturally associated distance, $d_f :\mathbf{R} \times \mathbf{R} \longrightarrow \mathbb{R}$, which is the absolute value of the difference between two attained values or most commonly known as the \emph{Euclidean} distance of them:
\begin{equation} \label{eq:euclidean-dist}
d_{f}(\mathbf{\hat{r}_1},\mathbf{\hat{r}_2})= \vert f(\mathbf{\hat{r}_1}) - f(\mathbf{\hat{r}_2}) \vert
\end{equation}
\begin{remark}
\label{rem:edge-weight}
By weighting each edge, $\mathbf{\hat{r}_1}\mathbf{\hat{r}_2}$, of the associated Hasse diagram, $G_{f}$, with the value $\vert f(\mathbf{\hat{r}_1}) - f(\mathbf{\hat{r}_2}) \vert$, it trivially holds that the distance $d_{f}$ between a pair of elements is the minimum length distance on this edge-weighted graph, i.e., $d_{f}$ is the natural distance on the Hasse diagram, $G_{f}$.
\end{remark}

Therefore, every IR evaluation measure, $f$, has a naturally associated context that is intrinsically derived from the measure itself. This context is composed of its associated ordering, $\preceq_{f}$, and distance, $d_{f}$. These two mathematical tools are inherently connected to the retrieval measure, since the ordering directly reflects the purpose for which it was designed, and the distance is the usual manner to perform comparisons on the attained values. They are intrinsic entities that only depend on the definition of the retrieval measure itself, and can be considered jointly to establish the following concept.

\begin{definition}
\label{def:intrinsic-framework}
The \emph{intrinsic framework} of an IR evaluation measure, $f$, is the set of possible rankings endowed with its associated ordering, $\preceq_f$, and its associated distance, $d_f$; it is denoted by $(\mathbf{R}, \preceq_f, d_f)$. 
\end{definition}

The intrinsic framework is closely related to the mathematical concept of \emph{intrinsic geometry}~\cite{do2016differential,gauss1828disquisitiones} of curves and surfaces\footnote{Imagine hypothetical beings living on the surface of a two-dimensional Euclidean space, $\mathbb{R}^2$, ignorant of the surrounding three-dimensional space (but with a sense of Euclidean distance). These beings are local observers, whose view reaches only a two coordinated environment. The geometrical elements of this surface capable of being observed or measured by these beings (essentially lengths) constitute what is called the \emph{intrinsic geometry} of the surface. The intrinsic properties of the surface are those which depend exclusively on the surface itself.}. Consider the reduction or quotient $(\mathbf{R^{*}}, \preceq_f^{*})$ of the ordering associated with a retrieval measure. As $\preceq_f^{*}$ is a total order, its associated Hasse diagram, $G_{f}^{*}$, is a chain that joins consecutive elements with straight lines. By weighting each edge as indicated in Remark~\ref{rem:edge-weight}, the Hasse diagram, $G_{f}^{*}$, can be considered a curve on its own, whose distance between any pair of contiguous elements is the associated weight. The intrinsic framework represents the intrinsic geometry of $G_{f}^{*}$, where the distance is computed with the minimum path length. In a strict sense, the intrinsic framework is not the intrinsic geometry of $G_{f}^{*}$ since it is a discrete curve, which lacks differentiability. However, this intrinsic framework enables to measure distances on $G_{f}^{*}$ with mathematical tools derived from the retrieval measure itself.

Comparisons on the intrinsic framework of a retrieval measure have sense since they are derived from the measure itself. If we consider a different retrieval measure, then the ordering of the intrinsic framework is changed. However, the criterion for measuring the minimum path length of the new measure will remain coherent since it verifies the axiomatic properties of a distance. Another property of the intrinsic framework is that it considers rankings of any size since no assumption has been made about that. In addition, no assumption has been made about relevance grades, then the intrinsic framework can handle binary or multi-graded relevance labels.

\section{Intrinsic Properties of an IR Evaluation Measure}
\label{sec:intrinsic-prop}
In this section, the metric and scale properties of the intrinsic framework are characterised.

\subsection{Metric Properties of an IR Evaluation Measure}
\label{subse:metric}
Let us recall some basic notions of metric spaces~\cite{frechet1906quelques,hausdorff2005set}.  Consider an IR evaluation measure, $f$; the associated distance, $d_{f}$, is a \emph{pseudometric} if it verifies the following two properties: (i) \emph{symmetry}, $d_f(\mathbf{\hat{r}_1},\mathbf{\hat{r}_2}) = d_f(\mathbf{\hat{r}_2}, \mathbf{\hat{r}_1})$, and (ii) \emph{triangular inequality}, $d_f(\mathbf{\hat{r}_1},\mathbf{\hat{r}_3}) \leq d_f(\mathbf{\hat{r}_1},\mathbf{\hat{r}_2}) + d_f(\mathbf{\hat{r}_2},\mathbf{\hat{r}_3})$. If $d_f$ also verifies (iii) the \emph{identity of indiscernible}, $d_f(\mathbf{\hat{r}_1},\mathbf{\hat{r}_2})=0 \Leftrightarrow \mathbf{\hat{r}_1}=\mathbf{\hat{r}_2}$, then $d_{f}$ is a metric. 

Strictly speaking it must be distinguished the evaluation measure, $f$, from the distance, $d_f$. According to the previous paragraph, the metric properties should be applied to the distance with statements such as ``this \emph{distance} is a pseudometric''. However, in the retrieval context, as the distance is derived from the IR evaluation measure, it is very common to assign the metric properties to the retrieval measure with statements such as ``this \emph{retrieval measure} is a metric''. The following result shows that every retrieval measure is a pseudometric.

\begin{proposition}
\label{prop:ir-mesure-isovaluation}
Let $(\mathbf{R}, \preceq_{f}, d_{f})$ be the intrinsic framework of an IR evaluation measure, $f$, then the associated distance, $d_{f}$, is a pseudometric.
\end{proposition}

In the retrieval scenario, it has been coined the term ``IR evaluation \emph{metric}'' to design any IR evaluation measure. However, the following result shows that not every retrieval measure is a metric. 
\begin{proposition}
\label{prop:caracter-metric}
Let $(\mathbf{R}, \preceq_{f}, d_{f})$ be the intrinsic framework of an IR evaluation measure, if $f$ is an injective or one-to-one function\footnote{In basic algebra~\cite{fraleigh2003first,hungerford2012algebra,jacobson2012basic}, $f$ is an injective function, if $f$ maps distinct elements to distinct elements, formally: $f(\mathbf{\hat{r}_1}) = f(\mathbf{\hat{r}_2})$ implies $\mathbf{\hat{r}_1} = \mathbf{\hat{r}_2}$, $\forall \mathbf{\hat{r}_1}$, $\mathbf{\hat{r}_2} \in \mathbf{R}$.}, then the associated distance, $d_f$, is a metric.
\end{proposition}

Thus, every retrieval measure is a pseudometric, and only the retrieval measures that assign distinct values to every system output ranking are metrics. 

\subsection{Scale Properties of an IR Evaluation Measure}
\label{subsec:ordinal-charact}
From the point of view of the RTM, a consistent assignment of real numbers to the empirical domain is a \emph{scale} for the attribute to be measured~\cite{stevens1951mathematics}. Stevens~\cite{stevens1946theory} distinguished four main types of measurement scales: \emph{nominal}, \emph{ordinal}, \emph{interval} and \emph{ratio}. By considering an order relationship, $\trianglelefteq$, on the underlying empirical domain; the scales, $\varphi$, reflecting or preserving this ordering are called \emph{ordinal scales}, formally: $a \trianglelefteq b \Leftrightarrow \varphi(a) \leq \varphi(b)$. 

In the retrieval scenario, every IR evaluation measure, $f$, defined on the ordering structure, $(\mathbf{R}, \preceq_{f})$, is an ordinal scale since the definition of the associated ordering (see Equation~\ref{eq:associted-ordering}) just verifies this property.

The intrinsic framework, $(\mathbf{R}, \preceq_{f}, d_{f})$, of a retrieval measure allows to consider an interval, defined as the set of elements between two end-points, formally: $[\mathbf{\hat{r}_1}, \mathbf{\hat{r}_2}] = \{\mathbf{\hat{r}} \in \mathbf{R} : \mathbf{\hat{r}_1} \preceq_f \mathbf{\hat{r}} \preceq_f \mathbf{\hat{r}_2}\}$. It can be quantified by its cardinality or \emph{span of the interval}, denoted by $\Delta_{\mathbf{\hat{r}_1} \mathbf{\hat{r}_2}}$, formally: $\Delta_{\mathbf{\hat{r}_1} \mathbf{\hat{r}_2}} = \vert [\mathbf{\hat{r}_1}, \mathbf{\hat{r}_2}] \vert$. The span of an interval represents how closely spaced is a pair of rankings. 

Then, an order relationship, defined on the set of possible intervals, can be introduced as follows: $[\mathbf{\hat{r}_1}, \mathbf{\hat{r}_2}] \preceq_{d_{f}} [\mathbf{\hat{r}_3}, \mathbf{\hat{r}_4}] \Leftrightarrow \Delta_{\mathbf{\hat{r}_1} \mathbf{\hat{r}_2}} \leq \Delta_{\mathbf{\hat{r}_3} \mathbf{\hat{r}_4}} $. This order relationship, $\preceq_{d_{f}}$, is a weak order since every pair of intervals is comparable and the transitivity is verified trivially. 

Following the RTM~\cite{roberts1985measurement}, an IR evaluation measure, $f$, defined on the ordering structure $(\mathbf{R}, \preceq_{f})$ is an \emph{interval scale} if it preserves differences, i.e., if equally spaced intervals are assigned to equal differences, formally:
\[
[\mathbf{\hat{r}_1}, \mathbf{\hat{r}_2}] \preceq_{d_f} [\mathbf{\hat{r}_3}, \mathbf{\hat{r}_4}] \Longleftrightarrow f(\mathbf{\hat{r}_2}) - f(\mathbf{\hat{r}_1}) \leq f(\mathbf{\hat{r}_4}) - f(\mathbf{\hat{r}_3}) \ .
\]

The following result characterise the interval scales.
\begin{proposition}
\label{prop:interval-constant-assi}
Consider an ordinal scale, $f$, which is a metric, then $f$ is an interval scale if and only if the attained values are equally spaced.
\end{proposition}

Thus, every retrieval measure is an ordinal scale on its associated ordering, and only the retrieval metrics whose attained values are equispaced are interval scales.

\section{Intrinsic Taxonomy of IR Evaluation Measures}
\label{sec:intrinsic-categories}
The characterisations of Section \ref{sec:intrinsic-prop} enables the classification of retrieval measures into three main categories, which provide a ready-to-use rule to identify intrinsic properties in terms of the attained values:
\begin{enumerate}
\item Every IR evaluation measure, $f$, is an ordinal scale and a pseudometric on its intrinsic framework. By default, these are the properties  of any retrieval measure. This category is denominated \textbf{ordinal/pseudometric}.
\item If the attained values of $f$ are different for every system output ranking, i.e., the retrieval measure is a one-to-one function, then $f$ is a metric (not necessarily an interval scale). These retrieval measures are \textbf{ordinal/metric}.
\item If the attained values of $f$ are equally spaced, then the retrieval measure is an interval scale. This category is denominated \textbf{interval/metric}.
\end{enumerate}

A quick glance at this taxonomy confirms that there are no ratio scales. The reason is that the intrinsic framework aims to deduce the metric and scale properties of an IR evaluation measure from the information contained in the retrieval measure itself, i.e., from its associated ordering and distance. Ratio scales need an additional operator among rankings~\cite{michell2014introduction,roberts1985measurement}, which is not present in the definition of the retrieval measure. Therefore, IR evaluation measures can be ratio scales, but an \emph{extrinsic} operation among system output rankings has to be previously specified.

\section{Some Examples}
\label{sec:examples}
In this section, some common retrieval measures are classified according to the taxonomy of Section \ref{sec:intrinsic-categories}. The empirical domain of each retrieval measure is its intrinsic framework\footnote{As noted in Section \ref{sec:intrinsic-prop}, the intrinsic properties of a retrieval measure deduced with this framework are based on the RTM.}.

\subsection{Set-Based Retrieval}
In this case, an IR system returns a set of documents in response to a query; by denoting with $1$ a relevant document and with $0$ a nonrelevant document, some examples of system outputs are as follows: $\mathbf{\hat{r}_1}=\{0, 0, 0, 0, 0\}$, $\mathbf{\hat{r}_2}=\{1, 0, 0, 0, 0\}$ and $\mathbf{\hat{r}_3}=\{1, 1, 0, 0, 0\}$. In general, the system output can be summarised with a contingency table of two binary variables: relevance and retrieval. Table~\ref{tab:contingency} illustrates the frequency distribution, where the two factors are shown simultaneously. 
\begin{table}
	\centering
  \caption{Contingency table of the set-based retrieval in the binary case.}
  \label{tab:contingency}
  \scalebox{0.75}
  {
  \begin{tabular}{c|c|c|}
    & \specialcell{\textbf{Relevant}\\ \textbf{documents}} & \specialcell{\textbf{Non relevant}\\ \textbf{documents}}\\
    \hline
    \specialcell{\textbf{Documents}\\ \textbf{retrieved}} & \specialcell{$tp$\\True Positive} & \specialcell{$fp$\\False Positive} \\
    \hline
    \specialcell{\textbf{Documents}\\ \textbf{non retrieved}} & \specialcell{$fn$\\False Negative} & \specialcell{$tn$\\True Negative} \\
    \hline
\end{tabular}
}
\end{table}

Some widely used evaluation measures are recall~\cite{van1979information}, precision~\cite{van1979information}, fallout~\cite{van1979information}, classification accuracy~\cite{belew2000finding}, miss rate, and error rate, defined as follows:
\begin{align*}
\textnormal{recall}&=\frac{tp}{(tp + fn)} &  \textnormal{precision} &=\frac{tp}{(tp + fp)} \\
\textnormal{fallout}&=\frac{fp}{(fp + tn)} & \textnormal{classification accuracy}&=\frac{(tp + tn)}{(tp + fn + fp + tn)} \\
\textnormal{miss rate}&=\frac{fn}{(tp + fn)}  &  \textnormal{error rate}&=\frac{(fp + fn)}{(tp + fn + fp + tn)} 
\end{align*}
All of them are metrics since the numerator of their analytical expressions has at least one of the terms: $tp$, $tn$, $fp$ or $fn$, which are distinct for every new relevant document present in the system output. In addition, they are interval scales since the denominator of their analytical expression is constant and the numerator increases one unit with every relevant retrieved document. Thus, their attained values are equispaced.

Other related evaluation measures are inverse recall, inverse precision, specificity, false discovery rate, false omission rate, and F-measure~\cite{van1979information}, defined as follows:
\begin{align*}
\textnormal{inverse recall}&=\frac{tn}{(fp + tn)} &  \textnormal{inverse precision} &=\frac{tn}{(fn + tn)} \\
\textnormal{specificity}&=\frac{tn}{(tn + fp)} & \textnormal{false discovery rate}&=\frac{fp}{(fp + tp)} \\
\textnormal{false omission rate}&=\frac{fn}{(fn + tn)}  &  \textnormal{F-measure}&=\frac{2 \cdot \textnormal{prec} \cdot \textnormal{recall}}{(\textnormal{prec} + \textnormal{recall})}
\end{align*}
The $F$-measure is a metric since it attains different values (the harmonic mean is an increasing function). The rest of the measures are metrics and interval scales since they verify the same properties indicated in the previous paragraph. However, the $F$-measure is not an interval scale; for instance, consider a collection of $15$ documents, where $5$ are relevant to a topic, then $F(\mathbf{\hat{r}_1}) = 0.000$, $F(\mathbf{\hat{r}_2})= 0.300$ and $F(\mathbf{\hat{r}_3}) = 0.509$. 

The generality factor or prevalence $= (tp + fn) / (tp + fn + fp + tn)$~\cite{van1979information} is a pseudometric since all its values are the same for every system output.

The utility measure~\cite{salton1983introduction}, $utility = \alpha \cdot tp + \beta \cdot fn + \gamma \cdot fp + \delta \cdot tn$, where $\alpha$, $\beta$, $\gamma$ and $\delta$ are the positive weights assigned by the user, present several possibilities varying the four parameters. In general, if the four parameters are different, then the utility measure is a metric, and some combinations of parameters could yield an interval scale.

Some user-oriented measures are based on the following four variables: (i) total number of relevant documents known to the user: $U$; (ii) number of relevant documents known to the user, which were retrieved: $R_k$; (iii) number of relevant documents unknown to the user, which were retrieved: $R_u$; and (iv) the number of retrieved documents: $A$. Some examples of user-oriented measures are coverage ratio~\cite{korfhageinformation}, retrieval recall~\cite{korfhageinformation}, novelty ratio~\cite{korfhageinformation}, and recall effort~\cite{korfhageinformation}, defined as follows:
\begin{align*}
\textnormal{coverage ratio}&=\frac{R_k}{U} &  \textnormal{retrieval recall} &=\frac{(R_k + R_u)}{U} \\
\textnormal{novelty ratio}&=\frac{R_u}{(R_u + R_k)} & \textnormal{recall effort}&=\frac{U}{A} 
\end{align*}
The relevance recall and the novelty ratio are not metrics by considering the following two system outputs: (i) $\mathbf{\hat{r}_5} =\{ 1$ retrieved relevant document known to the user$\}$; and (ii) $\mathbf{\hat{r}_6} = \{1$ retrieved relevant document known to the user, $1$ non-relevant document retrieved$\}$. The relevance recall and the novelty ratio attain the same value on these system outputs. The recall effort is neither a metric since it attains the same value on the following two system outputs: (i) $\mathbf{\hat{r}_7} =\{ 1$ retrieved relevant document unknown to the user, $1$ non-relevant document retrieved$\}$; and (ii) $\mathbf{\hat{r}_8} = \{2$ non-relevant documents retrieved$\}$. The coverage ratio is not a metric by considering two system outputs, which differ in one non-relevant retrieved document. 

Table~\ref{tab1} provides a summary of the intrinsic properties of these retrieval measures. 
\begin{table}
\centering
\caption{Intrinsic properties of some retrieval measures in the set-based retrieval.}\label{tab1}
\begin{tabular}{|l|c|c|c|}
\hline
 & \textbf{ord/pseudom} & \textbf{ord/metr} & \textbf{interv/metr} \\
\hline
recall~\cite{van1979information} & & & \checkmark \\
precision~\cite{van1979information} & & & \checkmark \\
fallout~\cite{van1979information} & & & \checkmark \\
miss rate & & & \checkmark \\
classification accuracy~\cite{belew2000finding} & & & \checkmark \\
error rate & & & \checkmark \\
inverse recall & & & \checkmark \\
inverse precision & & & \checkmark \\
specificity & & & \checkmark \\
false discovery rate & & & \checkmark \\
false Omission Rate & & & \checkmark \\
$F$-measure~\cite{van1979information} & & \checkmark & \\
generality factor~\cite{van1979information} & \checkmark & & \\
coverage ratio~\cite{korfhageinformation} & \checkmark & &\\
retrieval recall~\cite{korfhageinformation} & \checkmark & & \\
novelty ratio~\cite{korfhageinformation} & \checkmark & & \\
recall effort~\cite{korfhageinformation} & \checkmark & & \\
\hline
\end{tabular}
\end{table}

\subsection{Rank-Based Retrieval}
In this case, an IR system returns a ranked list of documents, $\mathbf{\hat{r}}= \langle d_1, d_2,\ldots \rangle$, in response to a query. Once documents have been retrieved, in the multi-graded case, the \emph{qrels} are used to determine the relevance of the retrieved documents to the topic. Denoting by $r$ a rank position of $\mathbf{\hat{r}}$, each retrieved document, $d_r$, is assigned a relevance degree, $\mathit{a}_r$, for $r\geq 1$. For instance, in the binary case, denoting with $\mathit{a}_1$ a relevant document and with $\mathit{a}_0$ a nonrelevant document, some example rankings of length four are as follows: $\mathbf{\hat{r}_1}=\langle\mathit{a}_1, \mathit{a}_0, \mathit{a}_0, \mathit{a}_0\rangle$, $\mathbf{\hat{r}_2}=\langle\mathit{a}_0, \mathit{a}_1, \mathit{a}_0, \mathit{a}_0\rangle$ and $\mathbf{\hat{r}_3}=\langle\mathit{a}_0, \mathit{a}_1, \mathit{a}_0, \mathit{a}_1\rangle$. These relevance degrees can be categorical labels, to handle numerical values, a \emph{gain function} is considered, $gain(\cdot)$, by assigning a positive real number to each relevance degree. The gain at rank $r$ will be denoted by $g(r) = gain(\mathit{a}_r)$, where $g(r)=0$ for a non-relevant document. 
For instance, in the binary case, if $g(\mathit{a}_1) = 1$ and $g(\mathit{a}_0) = 0$, then the previous ranking examples can be expressed as $\mathbf{\hat{r}_1}=\langle 1, 0, 0, 0\rangle$, $\mathbf{\hat{r}_2}=\langle 0, 1, 0, 0 \rangle$ and $\mathbf{\hat{r}_3}=\langle 0, 1, 0, 1 \rangle$. The cumulative gain at rank $r$ is the total gain up to rank $r$, which is defined as $cg(r) = g(r) + g(r-1)$, for $r >1$ and $cg(1)=g(1)$.

In this multi-graded context, the precision at rank $r$~\cite{van1979information,buttcher2007reliable,ferrante2018general} can be expressed as follows:
\[
\textnormal{Prec@}r = \frac{cg(r)}{r}\ ,
\]
This retrieval measure is a pseudometric, but not a metric since it holds that Prec@$4(\mathbf{\hat{r}_1}) = 0.250 =$  Prec@$4(\mathbf{\hat{r}_2})$. Similarly, the recall at rank $r$ is a pseudometric since its analytical expression is also based on the cumulative gain at $r$.

In this scenario, an \emph{ideal} ranking can be considered by listing first all documents with the higher relevance degree, then all documents with the contiguous relevance degree, and so on. Denoting by $ig(r)$ the gain at rank $r$ for an ideal ranking, $cig(r)$ the cumulative gain at rank $r$ for an ideal ranking, and by $L$ the length of a ranking, then the sliding ratio~\cite{pollock1968measures,korfhageinformation}:
\[
\textnormal{sr} = \frac{cg(L)}{cig(L)}\ ,
\]
is a pseudometric since sr$(\mathbf{\hat{r}_1}) = 1.000 =$ sr$(\mathbf{\hat{r}_2})$. A modified version of the sliding ratio~\cite{sagara2002performance}:
\[
\textnormal{msr} = \frac{\sum_{r=1}^L \frac{1}{r} g(r)}{\sum_{r=1}^L \frac{1}{r} ig(r)}\ ,
\]
is a metric since it is sensitive to the ranking order. However, it is not an interval scale since msr$(\mathbf{\hat{r}_1})=1$, msr$(\mathbf{\hat{r}_2})=0.5$ and msr$(\langle 0,0,1,0 \rangle) =0.333$. 

To present the following retrieval measures, we need to introduce the indicator function $isrel(r)$, defined as $isrel(r)=1$ if the document at rank $r$ is relevant, and $isrel(r)=0$ otherwise. Thus, the value $count(r) = \sum_{k=1}^r isrel(k)$ is the number of relevant documents within the top $r$ documents of the ranking.

Let $R$ be the total number of relevant retrieved documents, then the R-precision:
\[
\textnormal{R-precision} = \frac{count(R)}{R}\ ,
\]
is a pseudometric; considering the ranking $\mathbf{\hat{r}_4}=\langle 1, 0, 0, 1\rangle$, it holds that R-precision$(\mathbf{\hat{r}_3}) = 0.500 = $R-precision$(\mathbf{\hat{r}_4})$. 

The normalized recall~\cite{rocchio1964performance}:
\[
\textnormal{R\textsubscript{norm}} = 1 - \frac{\sum_{r=1}^R isrel(r) \cdot r - \sum_{r=1}^R r}{R \cdot (L-R)}\ ,
\]
is an interval scale in the binary case since every relevant document retrieved contributes one unity to the measure. 

The normalized precision~\cite{rocchio1964performance}:
\[
\textnormal{P\textsubscript{norm}} = 1 - \frac{\sum_{r=1}^R isrel(r) \cdot \ln r - \sum_{r=1}^R \ln r}{\ln \frac{L!}{R! \ (L-R)!}}\ ,
\]
is a metric, but not an interval scale. It verifies the same property than the normalized recall; however, the logarithm is not a constant increasing function.

The weighted R-precision~\cite{kando2001information}:
\[
\textnormal{R-WP} = \frac{cg(R)}{cig(R)}\ ,
\]
is a pseudometric since it attains the same values than the R-precision in the binary case. The R-measure~\cite{sakai2004new}, defined as:
\[
\textnormal{R-measure} = \frac{cg(R) + count(R)}{cig(R)+R}\ ,
\]
is a pseudometric since it attains the same values than the R-precision in the binary case. In the multi-graded case, the R-measure and R-WP are also pseudometrics since they attain the value $1$, for every ranking that all the top $R$ documents are (at least partially) relevant.

The average precision~\cite{buckley2017evaluating,hauff2010retrieval}:
\[
\textnormal{AP} = \frac{1}{R} \cdot \sum_{r = 1}^{L} isrel(r) \cdot \frac{count(r)}{r}\ ,
\]
is not a metric since it holds that AP$(\mathbf{\hat{r}_1}) = 0.250 =$ AP$(\mathbf{\hat{r}_3})$. As AP is not an interval scale, then the mean average precision on a set of $Q$ queries, MAP $= \frac{1}{Q} $ $ \sum_{i=1}^Q AP_i$, and the geometric mean average precision, GMAP $= \exp \frac{1}{Q}  \sum_{i=1}^{Q} $ $ \log AP_i$, should not be considered according to the permissible operations on the scale types~\cite{stevens1946theory} since they are means of ordinal values. This result confirms the findings of~\cite{robertson2006gmap}. The average weighted precision~\cite{kando2001information}:
\[
\textnormal{AWP} = \sum_{r = 1}^{L} isrel(r) \cdot \frac{cg(r)}{cig(r)}\ ,
\]
is not a metric. Considering the rankings $\mathbf{\hat{r}_1}$ and $\mathbf{\hat{r}_3}$, with only two relevant documents to the query, then AWP$(\mathbf{\hat{r}_1}) = 0.250 =$ AWP$(\mathbf{\hat{r}_3})$. 

The Q-measure~\cite{sakai2004new}, defined as:
\[
\textnormal{Q-measure} = \frac{1}{R} \cdot \sum_{r = 1}^{L} isrel(r) \cdot \frac{cg(r)+count(r)}{cig(r)+r}\ ,
\]
is neither a metric since Q-measure$(\mathbf{\hat{r}_1}) = 0.250 =$ Q-measure$(\mathbf{\hat{r}_3})$ (in the binary case, it attains the same values as AP). 

The reciprocal rank at rank $r$:
\[
\textnormal{RR\textsubscript{$r$}} = isrel(r) \cdot \frac{1}{r}\ ,
\]
is not a metric since RR\textsubscript{$4$}$(\mathbf{\hat{r}_2})= 0.500 =$ RR\textsubscript{$4$}$(\mathbf{\hat{r}_3})$. Thus, the expected reciprocal rank~\cite{chapelle2009expected,sirotkin2013search}:
\[
\textnormal{ERR} = \frac{1}{Q} \sum_{i=1}^Q \textnormal{RR\textsubscript{$i$}}\ ,
\]
should not be considered according to the permissible operations on the scale types~\cite{stevens1946theory} since it is a mean of ordinal values.

The discounted cumulative gain~\cite{kekalainen2002using,jarvelin2002cumulated}:
\[
\textnormal{DCG\textsubscript{$b$}} = \sum_{r = 1}^{L} \frac{g(r)}{\max \{1, \log_{b}r \}}\ , 
\]
is not a metric since DCG\textsubscript{$2$}$(\mathbf{\hat{r}_1}) = 1.000 =$ DCG\textsubscript{$2$}$(\mathbf{\hat{r}_2})$. 

The graded rank-biased precision~\cite{moffat2008rank,sakai2008information}:
\[
\textnormal{RBP\textsubscript{$p$}} = \frac{1 - p}{g(\mathit{a}_c)} \cdot \sum_{r = 1}^{L} p^{i-1} \cdot g(r)
\]
is not a metric, in general. For instance, consider the rankings $\langle 1, 0, 0 \rangle$ and $\langle 0, 1, 1\rangle$, the corresponding scores are $(1+0+0) / (1 - p)$ and $(0 + p + p^2) / (1 - p)$ respectively. Equating these expressions is obtained $1 = p + p^2$ with a real solution. Thus, for this parameter $p$, there are two rankings with the same score. Therefore, RBP\textsubscript{$p$} is not an interval scale. However, there are particular cases where it is an interval scale since their attained values are equispaced; for instance, RBP\textsubscript{$0.5$}~\cite{ferrante2021towards}.

The binary preference evaluation measure~\cite{buckley2004retrieval}:
\[
\textnormal{bpref} = \frac{1}{R} \cdot \sum_{r = 1}^L 1 - \frac{r-count(r)}{R}\ ,
\]
is a pseudo metric, but not a metric, since it attains the same value for a ranking with one relevant document in the first ranking position and a ranking with two relevant documents in the first and second ranking position.

An extension of the cumulated gain (CG) is the family of retrieval measures XCG~\cite{kazai2004report}. They consider the dependency of XML elements, such as overlap and near-misses. In this paper, an adapted version of these measures is considered, through the definition xG$[r] = g(r)$, for every ranking position, $r$. The user-oriented measure of normalised extended cumulated gain~\cite{kazai2005evaluation}:
\[
\textnormal{nxCG}[r] = \frac{xCG[i]}{xCI[i]} = \frac{cg(r)}{cig(i)}\ ,
\]
is not a metric since nxCG$[4](\mathbf{\hat{r}_1})= 1.000 =$ nxCG$[4](\mathbf{\hat{r}_2})$, when there is only one relevant document. Thus, the mean average nxCG at rank $r$~\cite{kazai2005evaluation}:
\[
\textnormal{MAnxCG}[r] = \frac{\sum_{j=1}^{r} \textnormal{nxCG}[j]}{r} = \frac{\sum_{j=1}^{r} \frac{cg(j)}{cig(j)}}{r}
\]
should not be considered according to the permissible operations on the scale types~\cite{stevens1946theory}. In addition, considering it as a measure, it is not a metric since MAnxCG$[4](\mathbf{\hat{r}_1}) = 0.250 =$ MAnxCG$[4](\mathbf{\hat{r}_2})$. 

The system-oriented effort-precision / gain-recall~\cite{kazai2005evaluation}:
\[
\textnormal{gr}[r] = \frac{\textnormal{xCG}[r]}{\textnormal{xCI}[L]} = \frac{cg(r)}{cig(n)}
\]
is not a metric since gr$[4](\mathbf{\hat{r}_1}) = 1.000 =$ gr$[4](\mathbf{\hat{r}_2})$.

Finally, the expected search length~\cite{cooper1968expected}, defined as:
\[
\textnormal{esl} = j + \frac{i \cdot s}{t+1}
\]
where $j$ is the total number of non-relevant documents in all levels preceding the final level; $t$ is the number of relevant documents in the final level; $i$ is the number of non-relevant documents in the final level and $s$ is the number of relevant documents required from the final level to satisfy the need according its type. The esl is not a metric in the Type 2 retrieval since it attains the same value for two rankings, which only differ in the order of the documents of a specific level.

In general, in the non-binary case of many retrieval measures, there are several ways in which different rankings can be awarded the same score. For instance, if the possible qrel values are: $\{\mathit{a}_0=0, \mathit{a}_1, \mathit{a}_2, \mathit{a}_3, \mathit{a}_4 = 1\}$, there are three variables to combine them in such a way that distinct rankings map to identical scores. Table~\ref{tab2} provides a summary of the intrinsic properties of these retrieval measures. 
\begin{table}
\centering
\caption{Intrinsic properties of some retrieval measures in the rank-based retrieval.}\label{tab2}
\begin{tabular}{|l|c|c|c|}
\hline
 & \textbf{ord/pseudom} & \textbf{ord/metr} & \textbf{interv/metr} \\
\hline
Prec@$r$~\cite{van1979information,buttcher2007reliable} & \checkmark & & \\
R-Precision & \checkmark & & \\
sliding ratio~\cite{korfhageinformation} & \checkmark & & \\
modified slid. ratio & & \checkmark & \\
R\textsubscript{norm}~\cite{rocchio1964performance} & & & \checkmark * \\
P\textsubscript{norm}~\cite{rocchio1964performance} & & \checkmark & \\
R-WP~\cite{kando2001information} & \checkmark & & \\
R-measure~\cite{sakai2004new} & \checkmark & & \\
Avg. Prec.~\cite{buckley2017evaluating,hauff2010retrieval} & \checkmark & & \\
AWP~\cite{kando2001information} & \checkmark & & \\
Q-measure~\cite{sakai2004new} & \checkmark & & \\
RR & \checkmark & & \\
DCG\textsubscript{$b$}~\cite{kekalainen2002using} & \checkmark & & \\
RBP\textsubscript{$p$}~\cite{moffat2008rank,sakai2008information} & \checkmark &  &  \\
bpref~\cite{buckley2004retrieval} & \checkmark & & \\
nxCG$[r]$~\cite{kazai2005evaluation} & \checkmark & & \\
MAnxCG$[r]$~\cite{kazai2005evaluation} & \checkmark & & \\
gr$[r]$~\cite{kazai2005evaluation} & \checkmark & & \\
esl~\cite{cooper1968expected} & \checkmark & & \\
\hline
\multicolumn{4}{l}{(*)\footnotesize{Only in the binary case.}} \\
\end{tabular}
\end{table}

\section{Conclusions}
\label{sec:conclusion}
As indicated in Section \ref{sec:introduction}, there are different approaches to determine the scale type of retrieval measures. The first step should be to theoretically ground their arguments, in order to make explicit the assumptions behind retrieval measures. Once these arguments have been correctly justified, then it is possible to assess whether one, both, or any other alternative are valid.

The results obtained here correspond to the representational paradigm, i.e., when the RTM is assumed. This paper has provided a theoretical basis of the metric and scale properties of a retrieval measure, when its empirical domain is not explicitly specified. These properties are deduced from the information contained in the retrieval measure itself, i.e., they are intrinsic properties. A taxonomy and a ready-to-use rule based on the attained values are introduced, and some common user-oriented and system-oriented retrieval measures have been classified according their intrinsic properties. It has been found that the strength of the set-based (first generation of) retrieval measures are their formal properties, most of them are metrics and interval scales. Thus, operations involving order, addition or difference operations among their attained values can be computed, according to the permissible operations on the scale types of Stevens~\cite{stevens1946theory}. On the other hand, rank-based (modern) retrieval measures attempt to capture more accurate aspects of systems' usefulness, dropping their formal properties. In general, they are pseudometrics and ordinal scales; thus, only operations involving the order of their attained values should be performed, according to the permissible operations. Thus, retrieval measures face a compromise between satisfying formal properties and capturing the user’s perception of usefulness.

The intrinsic framework can be useful to determine the metric and scale properties of emergent or existing IR evaluation measures, when the RTM is assumed. In addition, it enables to study other properties that exclusively depend on the retrieval measure itself, which is an interesting subject to be explored in future work. 

%
%
\bibliographystyle{spmpsci}
\bibliography{bibliography}

\appendix
\section{Appendix}
\subsection{Formal Proofs}
\begin{proof}
{\small
{\bf [Proposition \ref{prop:ir-mesure-isovaluation}]:} 

Symmetry is trivially verified since $d_{f}(\mathbf{\hat{r}_1},\mathbf{\hat{r}_2})= \vert f(\mathbf{\hat{r}_1}) - f(\mathbf{\hat{r}_2}) \vert = \vert f(\mathbf{\hat{r}_2}) - f(\mathbf{\hat{r}_1}) \vert = d_{f}(\mathbf{\hat{r}_2},\mathbf{\hat{r}_1})$. Triangular inequality is also trivial, by considering the triangular inequality on the real numbers: $\vert f(\mathbf{\hat{r}_1}) - f(\mathbf{\hat{r}_2}) \vert \leq \vert f(\mathbf{\hat{r}_1}) - f(\mathbf{\hat{r}_3}) \vert + \vert f(\mathbf{\hat{r}_3}) - f(\mathbf{\hat{r}_2}) \vert$.\qed 
}
\end{proof}

\begin{proof}
{\small
{\bf [Proposition \ref{prop:caracter-metric}]:} 

An interesting result about metric spaces~\cite{guccione2018espacios} states the following: ``\emph{Let} $(\mathbf{R_2}, d_2)$ \emph{be a metric space and let} $f:\mathbf{R_1} \longrightarrow \mathbf{R_2}$ \emph{an an injective or one-to-one function, then} $(\mathbf{R_1}, d_1)$ \emph{is a metric space, where} $d_1(\mathbf{\hat{r}_1}, \mathbf{\hat{r}_2}) = d_2(f(\mathbf{\hat{r}_1}), f(\mathbf{\hat{r}_2}))$, $\forall \mathbf{\hat{r}_1}$, $\mathbf{\hat{r}_2} \in \mathbf{R_1}$''.

In the retrieval scenario, $(\mathbf{R_2}, d_2) = (\mathbb{R}, \vert \cdot \vert)$, which is the metric space of the real line endowed with the usual norm (the absolute value). Let $f$ be a one-to-one IR evaluation measure; from the previous result, it follows that $(\mathbf{R_1}, d_1) = (\mathbf{R}, d_f)$ is a metric space, i.e., $d_f$ verifies the three postulates of a metric.\qed
}
\end{proof}

\begin{proof}
{\small
{\bf [Proposition \ref{prop:interval-constant-assi}]:} 

It will be seen the implication from right to left. Consider a metric ordinal scale, $f$, where the attained values are equally spaced.

An interval is called \emph{prime} if $[\mathbf{\hat{r}_1}, \mathbf{\hat{r}_2}] = \{\mathbf{\hat{r}_1}, \mathbf{\hat{r}_2}\}$. First, it will be seen that the function, $F(\mathbf{x}, \mathbf{y}) = \vert f(\mathbf{x}) - f(\mathbf{y}) \vert$, attains its minimum value on any prime interval. 

Let $[\mathbf{\hat{r}_1}, \mathbf{\hat{r}_3}] = \{\mathbf{\hat{r}_1}, \mathbf{\hat{r}_2}, \mathbf{\hat{r}_3}\}$ be a non-prime interval, where $\mathbf{\hat{r}_1} \preceq_f \mathbf{\hat{r}_2} \preceq_f \mathbf{\hat{r}_3}$, then it holds that $f(\mathbf{\hat{r}_1}) \leq f(\mathbf{\hat{r}_2}) \leq f(\mathbf{\hat{r}_3})$ since $f$ is an ordinal scale. It implies that $\vert f(\mathbf{\hat{r}_3}) - f(\mathbf{\hat{r}_1}) \vert \leq \vert f(\mathbf{\hat{r}_3}) - f(\mathbf{\hat{r}_2}) \vert + \vert f(\mathbf{\hat{r}_2}) - f(\mathbf{\hat{r}_1}) \vert$, i.e., the minimum value of $F$ is not attained at $[\mathbf{\hat{r}_1}, \mathbf{\hat{r}_3}]$. In addition, it holds that the function $F$ assign the same value for every prime interval. Given a prime interval, $[\mathbf{\hat{r}_1}, \mathbf{\hat{r}_2}]$, it can be considered one of its consecutive prime intervals, $[\mathbf{\hat{r}_2}, \mathbf{\hat{r}_3}]$, since $\preceq_f$ is a weak order (every pair of elements is comparable). These two prime intervals verify that $f(\mathbf{\hat{r}_1}) < f(\mathbf{\hat{r}_2}) < f(\mathbf{\hat{r}_3})$ since $f$ is a metric, and the attained values of $f$ are equally spaced. Thus, it can be assumed that $F(\mathbf{\hat{r}_1},\mathbf{\hat{r}_2}) = k \in \mathbb{R}^{+}$ for any prime interval $[\mathbf{\hat{r}_1}, \mathbf{\hat{r}_2}]$. 

Now, it will be seen that equally spaced intervals (not necessarily prime) are assigned equal differences. Consider any non-prime interval, $[\mathbf{\hat{r}_1}, \mathbf{\hat{r}_m}] = \{\mathbf{\hat{r}_1}, \mathbf{\hat{r}_2}, \ldots, \mathbf{\hat{r}_m}\}$. As $f$ is a metric, then it attains different values for different elements. Thus, it can be assumed that $f(\mathbf{\hat{r}_1}) < f(\mathbf{\hat{r}_2}) < \cdots < f(\mathbf{\hat{r}_{m-1}}) < f(\mathbf{\hat{r}_m})$. Then, every interval $[\mathbf{\hat{r}_i}, \mathbf{\hat{r}_{i+1}}]$ are prime intervals for $i=1, \ldots m-1$ since $F$ attain the minimum at these intervals. As $f(\mathbf{\hat{r}_m}) - f(\mathbf{\hat{r}_1}) = f(\mathbf{\hat{r}_m}) - f(\mathbf{\hat{r}_{m-1}}) + f(\mathbf{\hat{r}_{m-1}}) - \cdots - f(\mathbf{\hat{r}_2}) + f(\mathbf{\hat{r}_2}) - f(\mathbf{\hat{r}_1})$ and $f(\mathbf{\hat{r}_{i+1}}) - f(\mathbf{\hat{r}_i}) = k$ for $1 \leq i \leq m-1$, then $f(\mathbf{\hat{r}_1}) - f(\mathbf{\hat{r}_m}) = k \cdot m$, which only depends on the span of the interval, $m$, not on the considered elements. Therefore, equally spaced intervals are assigned equal differences, i.e., $f$ is an interval scale.

Finally, it will be seen the other implication. Consider any prime interval, $[\mathbf{\hat{r}_1}, \mathbf{\hat{r}_2}]$, of $\mathbf{R}$, as $f$ is an interval scale, then equally spaced intervals are assigned to equal differences, i.e., the value $\vert f(\mathbf{\hat{r}_2}) - f(\mathbf{\hat{r}_1}) \vert$ is constant for every prime interval of $\mathbf{R}$. In addition, it should be an strictly positive value. To see that the attained values are equally spaced, it is sufficient to check that different elements of $\mathbf{R}$ are assigned different values of $f$, which is hold since $f$ is a metric.\qed
}
\end{proof}
\end{document}